\documentclass[a4paper]{spie}  

\usepackage{amsmath,amsfonts,amssymb}
\usepackage{graphicx}
\usepackage[colorlinks=true, allcolors=blue]{hyperref}
\usepackage{enumitem}
\usepackage[dvipsnames]{xcolor}
\usepackage{multirow}
\usepackage{siunitx}

\def\pixelred{\color[RGB]{127, 9, 9}}
\def\pixelyellow{\color[RGB]{103, 102, 1}}
\def\pixelgreen{\color[RGB]{1, 102, 1}}
\def\pixelblue{\color[RGB]{3, 4, 120}}

\title{Overview of the Medium and High Frequency Telescopes of the LiteBIRD satellite mission}

\author[1]{L.~Montier}
\author[1]{B.~Mot}
\author[2]{P.~de Bernardis}
\author[3]{B.~Maffei}
\author[2,4]{G.~Pisano}
\author[2]{F.~Columbro}
\author[5]{J.E.~Gudmundsson}
\author[6]{S.~Henrot-Versillé}
\author[2]{L.~Lamagna}
\author[7]{J.~Montgomery}
% ...
\affil[1]{IRAP, Universit$\acute{\rm e}$ de Toulouse, CNRS, CNES, UPS, (Toulouse), France}
\affil[2]{Dipartimento di Fisica, Universit\`{a} La Sapienza, P. le A. Moro 2, Roma, Italy and INFN Roma}
\affil[3]{Institut d'Astrophysique Spatiale (IAS), CNRS, UMR 8617, Universit$\acute{\rm e}$ Paris-Sud 11, B$\hat{\rm a}$timent 121, 91405 Orsay, France}
\affil[4]{Cardiff University, School of Physics and Astronomy, Cardiff CF10 3XQ, UK}
\affil[5]{Stockholm University}
\affil[6]{Universit\'e Paris-Saclay, CNRS/IN2P3, IJCLab, 91405 Orsay, France}
\affil[7]{McGill University, Physics Department, Montreal, QC H3A 0G4, Canada}
% ...

\authorinfo{Further author information: L.~Montier (E-mail: ludovic.montier@irap.omp.eu)}

\begin{document} 
\maketitle
\begin{abstract}
LiteBIRD is a JAXA-led Strategic Large-Class mission designed to search for the existence of the primordial gravitational waves produced during the inflationary phase of the Universe, through the measurements of their imprint onto the polarization of the cosmic microwave background (CMB). These measurements, requiring unprecedented sensitivity,  will be performed over the full sky, at large angular scales, and over 15 frequency bands from 34\,GHz to 448\,GHz. The LiteBIRD instruments consist of three telescopes, namely the Low-, Medium- and High-Frequency Telescope (respectively LFT, MFT and HFT). We present in this paper an overview of the design of the Medium-Frequency Telescope (89--224\,GHz) and the High-Frequency Telescope (166--448\,GHz), the so-called MHFT, under European responsibility, which are two cryogenic refractive telescopes cooled down to 5\,K. They include a continuous rotating half-wave plate as the first optical element, two high-density polyethylene (HDPE) lenses and more than three thousand transition-edge sensor (TES) detectors cooled to 100\,mK. 
We provide an overview of the concept design and the remaining specific challenges that we have to face in order to achieve the scientific goals of LiteBIRD.
\end{abstract}


\section{INTRODUCTION}
\label{s:intro}

LiteBIRD, the Lite (Light) satellite for the study of $B$-mode polarization and Inflation from cosmic background Radiation Detection \cite{Hazumi:2019ltd, Sekimoto2018}, is a JAXA-led Strategic Large-Class mission, selected by ISAS/JAXA in 2019 to be launched by the end of the 2020s, and aimed at mapping the cosmic microwave background (CMB) polarized emission over the full sky at large angular scales. As the fourth generation of CMB space missions, after Planck, LiteBIRD is targeting the measurement of the CMB $B$-mode signals, which are known to be the best probe of the primordial gravitational waves generated during the first period of our Universe's history, as predicted by the cosmological inflation theory. These $B$ modes are large-scale curl patterns imprinted in the CMB by the primordial gravitational waves, and  characterized by a power spectrum whose amplitude is directly proportional to the tensor-to-scalar ratio (called $r$), which is related to the inflationary energy scale. Their detection would allow us to test major inflationary models and directly access the energy of inflation. While the current upper limit is $ r < 0.044$ at 95\,\% confidence level \cite{Tristram2020}, the mission goal of LiteBIRD is to measure $r$ with a precision of $ \delta r < 0.001$, including statistical errors, foregrounds contamination, systematics uncertainties, and margins. This will provide a crucial test of the cosmic inflation theory. 

LiteBIRD has been endorsed as one of the prioritized projects in the Master Plan 2020 of the Science Council of Japan. The project is currently supported in Phase~A by many other partners, including the U.S., Canada, France, Italy and Spain. The concept design has been studied by researchers from Japan, the U.S., Canada, and Europe since September 2016. 

The BICEP/Keck experiment \cite{bicep2planck} gave a lesson to the community about the importance of characterizing dusty Galactic foregrounds as contaminants for CMB polarization $B$-mode measurements with a limited range of frequencies. Hence, the frequency coverage of LiteBIRD has been extended to higher frequency bands, up to 448\,GHz, compared to the original design, which was 60--280\,GHz. At the same time, the collaboration has been opened to European expertise and Planck heritage to take charge of the development of the high-frequency channels to be included in the payload module. Currently, the Medium- and High-Frequency Telescopes (MHFT) cover observational frequencies from 89 to 448\,GHz, to achieve the science goals. A feasibility study to employ this frequency range was carried out in the framework of a Concept Design Facility (CDF) study at ESA in 2018.

The challenging scientific requirements of LiteBIRD imply stringent technical requirements to reach an unprecedented control of the instrumental systematic effects, especially since LiteBIRD mainly targets the largest scales over the sky (multipoles $2 < \ell < 200$). This requires a high sensitivity and strong mitigation of the $1/f$ noise, which led us to consider for all telescopes the use of a continuously rotating half-wave plate. While this technology has now been used more commonly on ground-based and balloon-borne CMB experiments, the adaptation to space represents a real challenge that we will describe here. Recent progresses made on detector arrays for ground-based and balloon-borne CMB experiments are also largely re-invested in the LiteBIRD design, while again strong efforts have been made to adapt such technologies to the space environment. 

We stress that the LFT, MFT and HFT are parts of a consistent setup of telescopes targeting a unique scientific goal, and considered as parts of the same LiteBIRD instrument. Hence the optimization process of the design has been done simultaneously on the whole frequency range, taking care of the band overlap between telescopes. While an overview of the LiteBIRD mission can be found in the companion SPIE paper \cite{Hazumi2020}, and  an  overview of the LFT is presented in the second companion paper \cite{Sekimoto2020}, we focus here on the MFT and HFT only.


\section{LiteBIRD MHFT concept}
\label{s:concept}

\subsection{Specifications}
\label{ss:requirements}

The technical requirements of the MHFT are derived from the scientific high-level requirements on the tensor-to-scalar ratio $r$, i.e., a total uncertainty $\delta r < 0.001$. This total error budget is distributed equivalently over three main contributions at a level of $5.7\times10^{-4}$ each: statistical uncertainty; systematic uncertainty; and margin. The statistical uncertainty requirement is strongly driven by the global sensitivity of the instrument, and also implies stringent constraints on the $1/f$ noise component and mitigation of the foregrounds contaminants. The systematics uncertainty budget is driven by the level of knowledge of the instruments. An uncertainty on the knowledge of a given instrumental parameter can be propagated to the inference on the tensor-to-scalar ratio $r$ and later translated as a systematic error $\sigma_{\rm syst}(r)$. The 
goals are to keep the uncertainty on $r$ smaller than $5.7\times10^{-6}$ for each source of instrumental systematic in the first iteration of the error budget distribution. 

We stress that such requirement flow-down analyses have been performed through global studies combining all LFT, MHFT, and HFT telescopes, considered as a global LiteBIRD instrument. Hence the information provided below for the main technical aspects is derived from this process and shared with\cite{Sekimoto2020}. These should be considered as current specifications or technical goals, related to high-level requirements, but not as formal technical requirements at this point. The numbers may change in the context of a further global and iterative optimization between LFT, MFT and HFT, during the concept study phase.  

\noindent
\textbf{\textit{Frequency coverage.}} The MFT and HFT telescopes are expected to cover the frequency range 89--448\,GHz, which allows us to characterise the Galactic dust component's polarized emission at higher frequencies and to increase sensitivity in the CMB bands.  

\noindent
\textbf{\textit{Band sensitivities.}} MHFT sensitivities per band have been optimized with LFT to reach requirements on $r$ and mitigate the contamination by Galactic Foregrounds, leading to the sensitivities of Table~\ref{tbl:sensitivities}. This optimization was performed within the so-called ``multipatch'' component-separation framework.~\cite{ErrardStompor2019} It assumes the presence of polarized CMB, dust and synchrotron signals, and considers spatial variability of foreground spectra, typically on $\sim 7^{\circ}$ angular scales. The optimization process tunes the sensitivity per band in order to reduce the foreground residuals in the recovered CMB map, while not increasing the noise and therefore lowering the overall error on $r$. We stress that this cannot be considered as a strong requirement of the sensitivity per band, but only a global sensitivity requirement integrated over all LFT, MFT and HFT bands, which can be found in Hazumi et al.\cite{Hazumi2020}

\noindent
{\textbf{\textit{$1/f$ noise.}}} The knee frequency of the post-demodulation $1/f$ noise should be below 0.1\,mHz (assuming a scanning strategy defined with 0.05\,rpm spin rate, $\alpha$=$45^\circ$ and $\beta$=$50^\circ$). The knee frequency of the raw $1/f$ noise should be well below 2.6\,Hz (39\,rpm$\times$4), or still to be defined for gain calibration. 

\noindent
\textbf{\textit{Data loss and operational duty cycle.}} The operating life of instruments should be long enough to perform observations for 3 years. The instrument should have an operational duty cycle of 85\,\% for science observations, including all downtime for cryogenic cycling, detector operation preparation, and data transfer. 

\noindent
\textbf{\textit{Angular resolution.}} The angular resolution of each detector's response shall be sufficient to cover the required $\ell$ range, i.e., $2 \le \ell \le 200$. It
should have a FWHM of 80 acrmin or better. 

\noindent
\textbf{\textit{Beams.}} The more stringent constraints on knowledge of the beam come from preliminary analyses at 100\,GHz. \textit{Near sidelobe} knowledge (up to $10^\circ$ from the beam peak) is expected to be known at the precision level of $-$30\,dB. Also, its beam pattern should be confirmed to be consistent with its designed pattern at a precision level of 10\,\% or better. \textit{Far sidelobes} (located above 0.2 rad) shall be known at the precision level of $-$56\,dB. These numbers are currently being refined with new dedicated analyses for the MFT and HFT. 

\noindent
\textbf{\textit{Thermal and structural requirements.}}  According to systems design, heat dissipation of MHFT is limited to 4\,mW, which includes polarization-modulator units and temperature control of MHFT optical components. The minimum eigen-frequency for MHFT is 100\,Hz and 50\,Hz for axial and lateral axes, respectively.

\subsection{Design rationales}
\label{ss:rationales}

The concept of LiteBIRD has been fully driven by the stringent scientific requirements, which proposes to focus on the largest scales over the sky of the polarization signal of the CMB to perform a combined detection of the reionization and recombination bumps of the $B$-mode spectrum due to primordial gravitational waves. Such a requirement implies reaching an extremely low level of $1/f$ noise, combined with high sensitivity. 

The first important design optimization triggered by the requirements above has been to integrate as the first optical element a continuously-rotating half-wave plate (HWP) for all telescopes. The presence of this continuously-rotating HWP performs an effective suppression of the $1/f$ noise, allowing us to distinguish between the instrumental polarized signal and the sky signal, which is modulated at $4f_{\rm HWP}$. A detailed trade-off analysis, including the polarization effects induced by the HWP itself, has been carried out between the two cases, i.e., with and without the HWP. It has been shown that the performance is degraded by a factor of about 2 without the HWP, and scaling as the ratio of $f_{\mathrm{knee}}$ over the spin rate of the satellite. Hence both MFT and HFT are equipped with a levitating continuously-rotating HWP, to minimize the heat load impact (see Sect.~\ref{ss:HWPM}).

The frequency coverage of the MHFT has been optimized simultaneously with the frequency coverage of the LFT, based on the constraints imposed by HWP materials. A minimum overlap between the bands has been included, in order to maximize sensitivity first, and allow cross-analysis of the systematics at the same frequency. The extension to higher frequencies, up to 448\,GHz was motivated due to the strong impact of Galactic foregrounds on CMB $B$-mode analysis\cite{bicep2planck}. The MFT finally covers 5 bands from 89\,GHz to 224\,GHz, while the HFT covers 4 bands from 166\,GHz to 448\,GHz.

The split of the 89-448\,GHz frequency range into two telescopes has been motivated by a trade-off analysis between a reflective versus refractive option, initiated in the framework of the ESA CDF study in March 2018. The two following designs have been studied: (i)
\textit{fully reflective}, a single reflective optics + reflective HWP covering the full frequency range; and (ii) \textit{fully refractive}, two telescopes with refractive optics + transmissive HWP, by splitting the total frequency range into mid-frequency (MFT) and high-frequency (HFT) ranges.

\begin{figure} [ht]
   \begin{center}
   \begin{tabular}{c} 
   \includegraphics[width=0.95\textwidth]{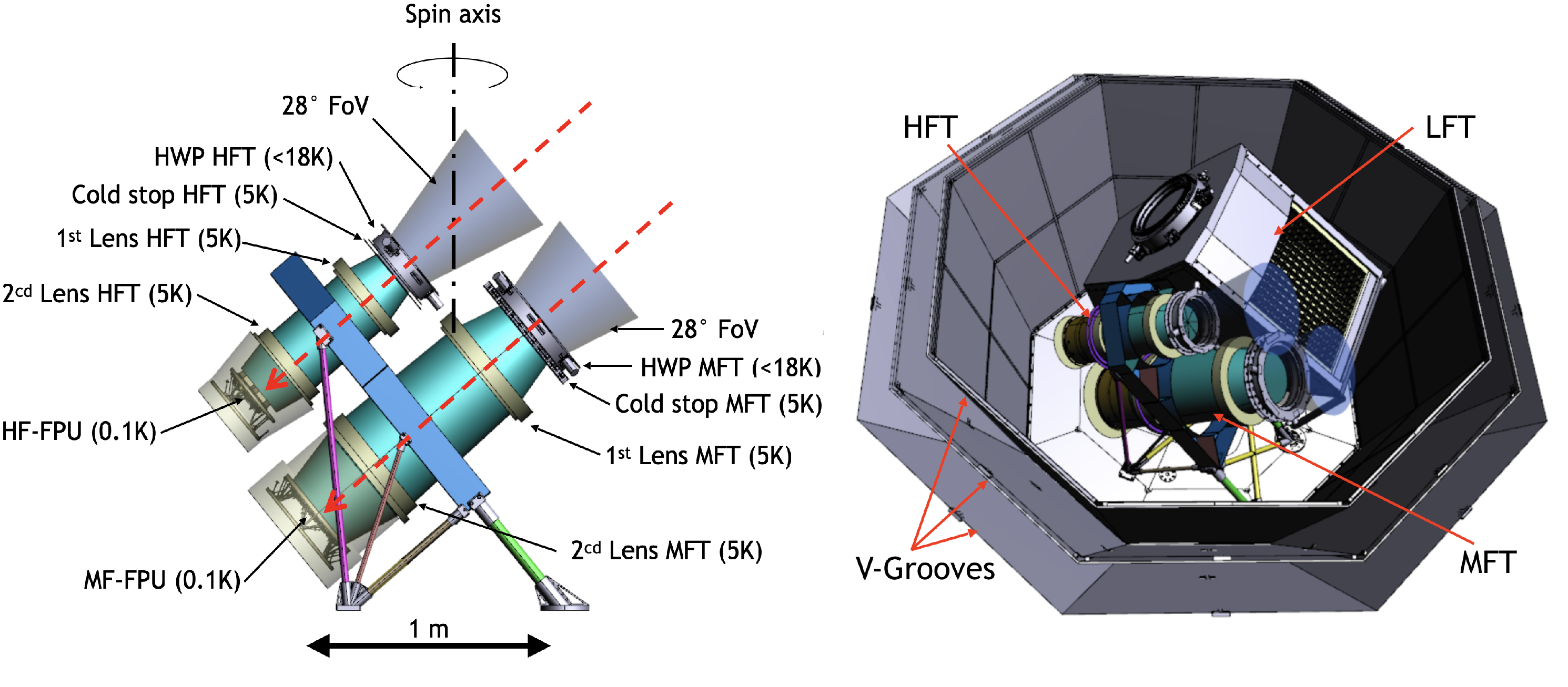}
   \end{tabular}
   \end{center}
   \caption[example] 
   { \label{fig:Overview} 
Left: MHFT overview. The various sub-systems comprising the telescopes are identified, and held by the mechanical structure. The optical front baffles still have to be designed. Right: overview showing the Low-, Medium- and High-Frequency Telescopes (respectively LFT, MFT and HFT) installed in the Payload Module.}
   \end{figure}

Further work and optimization has been performed on both design options after the ESA-CDF, and the major issues identified during the CDF have been fixed.
The main system level trade-off between reflective and refractive telescope has been done and allowed us to define the fully refractive option as the baseline for the MHFT (see left panel of  Fig.~\ref{fig:Overview}). It gives a very simple and compact design that takes advantage of strong heritage from ground-based and balloon-borne experiments, such as BICEP2\cite{BICEP2}, Keck\cite{KECK}, SPIDER\cite{SPIDER}, LSPE\cite{LSPE}, and Simons Observatory\cite{SimonsObs}. Furthermore, the split between middle and high frequencies facilitates the design of the filtering scheme and the calibration strategy.

\begin{table}[ht]
\centering
\begin{tabular}{|c|c|c|c|c|c|c|c|}
\hline
  \multicolumn{2}{|c|}{} & \multicolumn{3}{c|}{MFT} & \multicolumn{3}{c|}{HFT}   \\ \hline
 Band   & $\Delta\nu (GHz)$ & Beam size & NET array          & Sensitivity & Beam size & NET array  & Sensitivity   \\ 
 
  [${\rm GHz}$]  & [$\Delta\nu/\nu$] & [${\rm arcmin}$]  & [$\mu{\rm K} \sqrt{\rm s}$] & [$\mu{\rm K}\cdot{\rm arcmin}$] & [${\rm arcmin}$]  & [$\mu{\rm K} \sqrt{\rm s}$] & [$\mu{\rm K}\cdot{\rm arcmin}$] \\ 
\hline
100     &  23 (0.23)        &	37.8    &   4.19             &	8.47        &            &           &   \\  
119     &  36 (0.30)        &	33.6    &   2.82             &	5.69        &            &           &   \\ 
140     &  42 (0.30)        &	30.8    &	3.16             & 	6.39        &            &           &   \\
166     &  50 (0.30)        &	28.9    &   2.75             &	5.57        &            &           &    \\ 
195     &  59 (0.30)        &   28.0    &   3.48             &  7.03        & 28.6      &  5.19     &   10.49\\  
235     &	71 (0.30)       &	        &                    &              & 24.7	    &   5.34    &	10.79 \\  
280     &	84 (0.30)       &	        &                    &              &  22.5     &	6.82    &	13.80 \\ 
337     &	101 (0.30)      &	        &                    &              &	20.9    &	10.85   &	21.95 \\ 
402     &	92 (0.23)       &	        &                    &              & 17.9      &	23.45   &	47.45 \\   \hline
\end{tabular}
\caption{LiteBIRD MFT and HFT sensitivities. We stress that the overall sensitivity performance per band is obtained by combining these values with those of the LFT, which has common bands at 100, 119, and 140\,GHz. See Ref.~\citenum{Hazumi2020} for a complete sensitivity description.}
\label{tbl:sensitivities}
\end{table}

\section{LiteBIRD MHFT design}
\label{s:design}

We present below the current baseline design of the MFT and HFT, which both share the same optical and mechanical designs.

\subsection{Optical design}
\begin{figure}[t]
    \centering
    \begin{tabular}{cc}
    \begin{minipage}{0.5\hsize}
    \includegraphics[width = \hsize ]{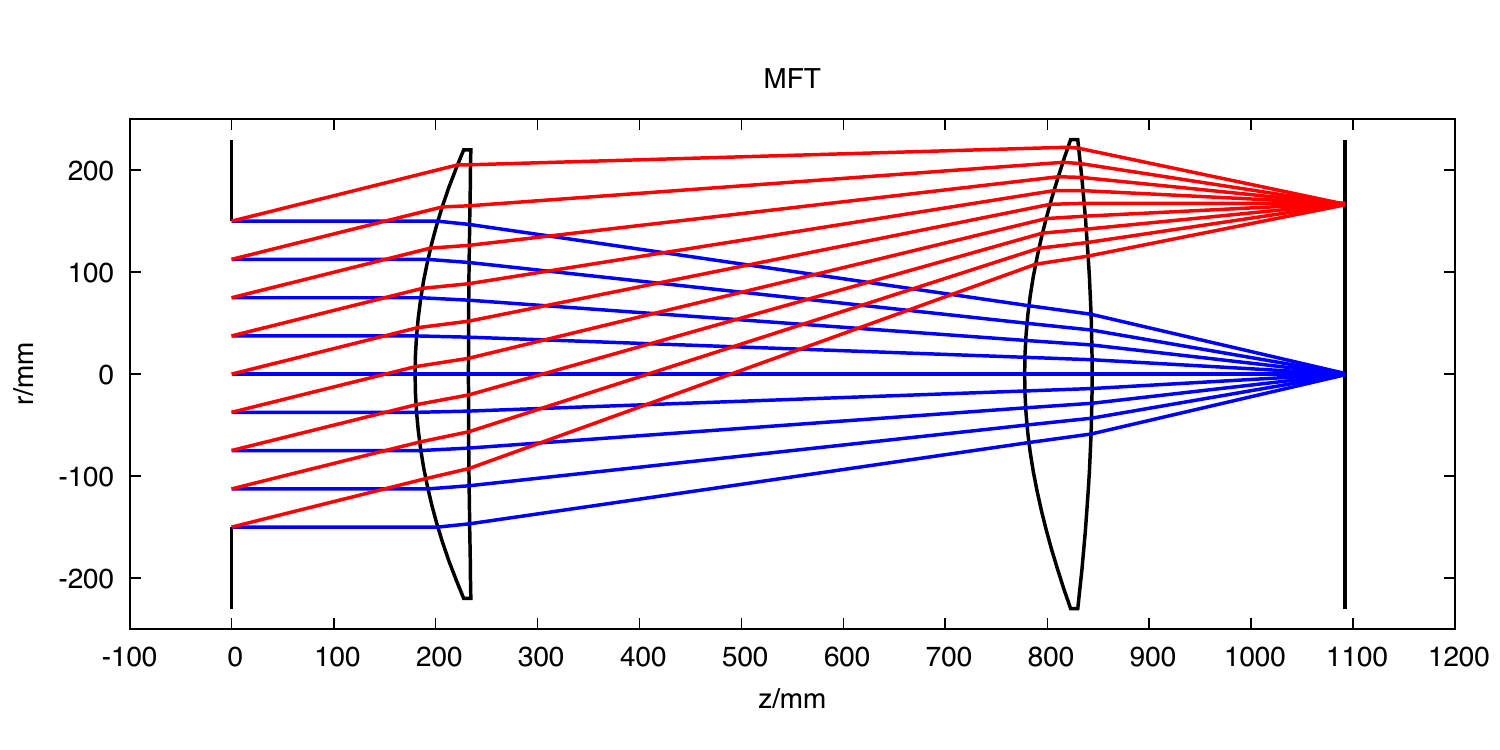}
    \end{minipage} &
    \begin{minipage}{0.5\hsize}
    \includegraphics[width = \hsize ]{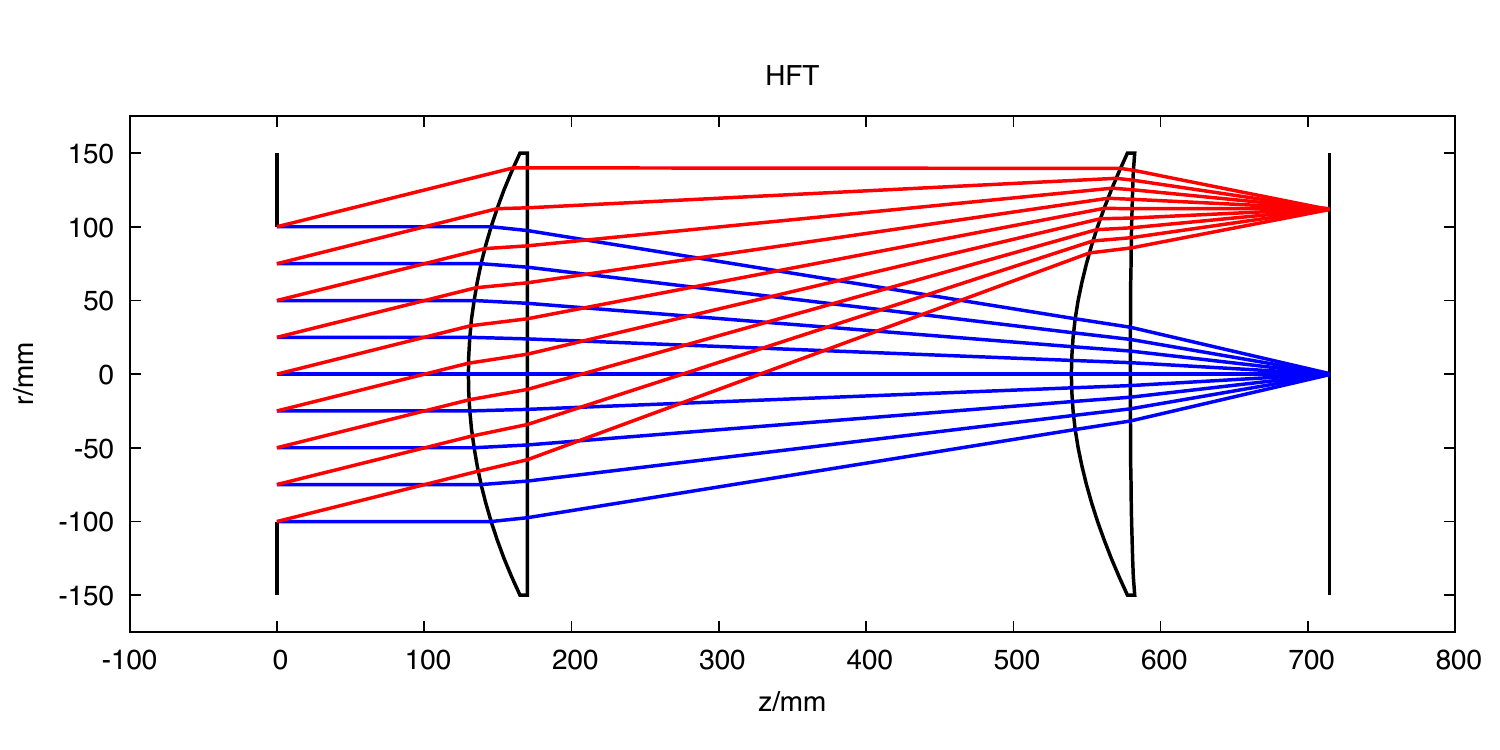}
    \end{minipage}
    \end{tabular}
    \caption{Ray diagrams of MFT (left) and HFT (right). The on-axis and off-axis fields (14$^\circ$) are the blue and red rays, respectively. The telescope aperture is located at $ z = 0 $.}
    \label{fig:opt_config}
\end{figure}

The current baseline MHFT employs two telescope barrels. The optical configurations are shown in Fig.~\ref{fig:opt_config}. Both telescopes feature two plastic lenses with assumed index of refraction $n_{\rm r}=1.52$. As further discussed in Sect.~\ref{sss:Lenses}, the current baseline assumes polypropylene (PP) for the plastics.
The actual choice of the material (ultra-high molecular weight polyethylene being an alternative to PP) affects the lens profiles marginally, and will be driven by further considerations and trade-off analysis. For example, dielectric losses at high frequency may become a discriminating parameter for the HFT, with PP being slightly disfavoured in spite of its higher melting point, a generally preferable feature for the deposition of broadband anti-reflection coatings. 

The two-lens designs of MFT and HFT are fully telecentric, with a maximum chief-ray incidence on the focal plane of \ang{0.1}. They feature an aperture stop of 300\,mm (MFT) and 200\,mm (HFT), located skywards of the objective lens. The diffraction-limited field of view is $28^{\circ}$, with a working f-number of 2.2. Basic analysis of the two systems has shown diffraction-limited performance across the whole spectral range up to the edge of the focal plane 
and a working f-number increase to $\sim$2.45 while moving to the edges of the field of view.

Both systems are fairly robust in terms of rotation and translation tolerance of the lenses and of their placement with respect to the apertures. A comparatively higher sensitivity to percent-level variations of the refractive index has been highlighted, mandating a careful characterization of the plastics at the operating temperature. The transmissive metal-mesh HWPs are accommodated in close proximity to the aperture stop in each telescope. Quasi-optical filters are placed along the optical chain, and absorbers on the tube walls and around the aperture stop are implemented to ensure further control of stray radiation. The telescope is cooled down to 4.8\,K. A 1.8-K cold hood surrounds the 100-mK Focal Plane Unit (FPU). Forebaffles at the sky input ensure control of beam sidelobes by preventing pickup from the payload structure close to each telescope.

A set of preliminary physical optics simulations, using simulation methodology described in Ref.~\citenum{Gudmundsson2020}, for this configuration have been performed to inform the first runs of time-domain simulations for the MHFT. These simulations provide estimates of main-beam far-field responses as well as spillover efficiencies for sensitivity calculations\cite{lamagna2021}. More advanced simulations are underway to capture a wider range of effects from the non-ideal optics chain. These include forebaffles, optics tubes, filters, imperfect rotating HWPs, dielectric losses, non-ideal coatings, absorbing materials, scattering off surfaces, profiled surface deviations, optical ghosts, and realistic feeds.
A laboratory testing effort (bread-board modeling) is underway to provide validation and highlight optical characterization criticalities for both telescopes.

\subsection{Quasi-optical components}
\label{ss:QO-Optics} 

\subsubsection{Lenses}
\label{sss:Lenses} 

The MHFT optical design employs lenses manufactured from polypropylene (PP), and anti-reflection coatings (ARCs) with matching layers of porous poly-tetrafluouroethane (pPTFE). This coated lens technology has been well characterized and validated through an extensive ESA-funded Technical Research Programme (TRP) programme. In addition, such lenses, of similar sizes, have been manufactured by members of the project team, and successfully deployed on ground-based and balloon-borne CMB polarimetry experiments, such as POLARBEAR\cite{Polarbear:article}, BICEP\cite{bicep}, BICEP2\cite{BICEP2}, SPIDER\cite{SPIDER}, ACT\cite{ACT}, Advanced ACT\cite{AdvancedActFieldedPerformance}, EBEX\cite{ebex} etc. In addition, porous PTFE coated lenses have also been deployed in space on the Herschel-SPIRE\cite{Herschel} satellite instrument.
As such, the coated lenses demonstrate a high-level of technical maturity, operating in a cryogenic vacuum environment. 

We note that the large relative bandwidths of the MFT and HFT instruments (2.5:1 and 2.7:1, respectively) require the use of multiple-layer coatings that would need additional materials with intermediate refractive indices. However, multi-octave anti-reflection coatings can still be realised using interspaced layers of the same PP and pPTFE materials, as demonstrated elsewhere\cite{Pisano2018a}. This approach is the baseline of our current ARC developments, which are part of an ongoing ESA TRP programme, where we also plan to accurately characterize these materials at cryogenic temperatures (including losses, stress-induced birefringence, etc.).  

Although conventional dielectric lenses are the baseline for the HFT, two new types of metamaterial lens, based on mesh-filter technology, have recently been designed, manufactured and characterized at millimetre wavelengths by members of the LiteBIRD EU team. \cite{Moseley2018}\cite{Pisano2018b} The first type is a graded-index lens\cite{Savini:12} based on standard gradient parabolic profiles, as well as Fresnel-type lenses. The second type is based on variable phase-delays changing across the planar surface of the device\cite{Pisano:13}. Both demonstration lenses mimic the behavior of 300-mm diameter f/3.5 and f/4 classical lenses and reproduce fairly accurate Airy patterns. The performance of these lenses has been characterized in terms of frequency dependence, focal point variability, and deviations from the Airy pattern due to diffraction effects. Other performance parameters and systematic effects currently under investigation are the off-axis performance, the cross-polarization and the transmission losses. Depending on the required diameter (300\,mm) and operational bandwidth ($\sim$2:1 to 3:1) metamaterial mesh-lenses can be considered as options for the MFT and HFT instruments, should we have sufficient confidence in their performance on the required timescale. The manufacturing processes and materials of the mesh-lenses are identical to those used for mesh-filters (which are TRL-9) and mesh-HWPs. In addition, these lenses are very thin (few mm), flat, robust, very light (few hundred grams) and have no issue with cryogenic operation.

\subsubsection{Half-wave plates }
\label{sss:HWP}

The current baseline design assumes to use of transmissive mesh HWP\cite{Pisano:12,Pisano:20} (M-HWP) technology for the two MFT and HFT polarization modulator units. 
Mesh HWPs have been developed to overcome the issues faced with sapphire HWPs (S-HWPs), such as the weight or (marginally) the realization of broadband ARCs. Using the well established mesh-technology, it is possible to design anisotropic filters able to arbitrarily phase-shift the radiation in orthogonal polarization directions. Mesh HWPs are stacks of copper grids embedded in a single robust polypropylene slab. These devices are very light, thin, can be easily AR-coated and can be cryogenically cooled. The limitation of these devices is their bandwidth, which currently cannot exceed the 3:1 range without adding substantial conductive losses. The current R\&D is focused on the realisation of low-loss grids, which will allow us to achieve larger bandwidths. M-HWPs for MFT and HFT working across bandwidths of the order of 2.5:1 to 2.7:1 can be currently implemented and realised with the required diameters, i.e., 300\,mm and 200\,mm, respectively. The current absorption losses -- due to the combined effect of the finite electrical conductivity of Copper and the dielectric losses of the polypropylene -- would be 1--3\,\%, although the current R\&D on low-loss grids is targeting their reduction down to the 1\,\% level.

We notice that in the context of the concept trade-off analysis performed in 2018, one more option has been studied for the reflective design, namely the embedded reflective HWP \cite{Pisano:16} (ER-HWP). 


\subsubsection{Filters}
\label{sss:Filters}

Although band definition is achieved via on-chip filtering, additional optical filters are required in order to control the out-of-band rejection level, to protect the detectors from stray light and to control the thermal environment.  For MFT we are considering deploying a chain of four low-pass filters positioned at the 4.8-K, 2-K, 300-mK and 100-mK stages.  It is noted that the HWP also acts as a low-pass filter and we have the option of adding a high-pass element at the detectors should this be required.  The average in-band transmission over the frequency range of the MFT (or HFT) for each of these elements can be considered to be 95\,\%.

\subsection{Polarization modulator units}
\label{ss:HWPM}

One of the key sub-systems of the LiteBIRD instrumental design is the continuously rotating HWP, which will permit much better control of some of the systematic effects\cite{Systematics:2016, Columbro2019}, in particular, alleviating the impact of the very low-frequency noise component on the polarization signal. In order to minimize the thermal impact of such a rotating mechanism, it has been chosen to adopt a magnetic levitating rotor\cite{Polarbear:article, Johnson2017}. One polarization modulation system is designed for each of the telescopes\cite{Columbro:2021}.

\begin{figure} [ht]
   \begin{center}
   \begin{tabular}{c} 
   \includegraphics[height=5cm]{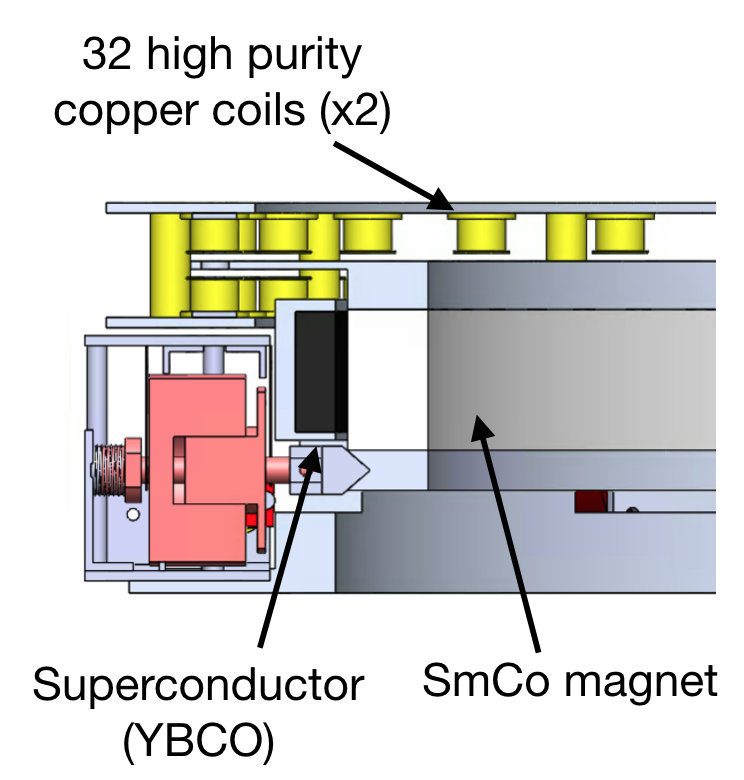}
      \includegraphics[height=5cm]{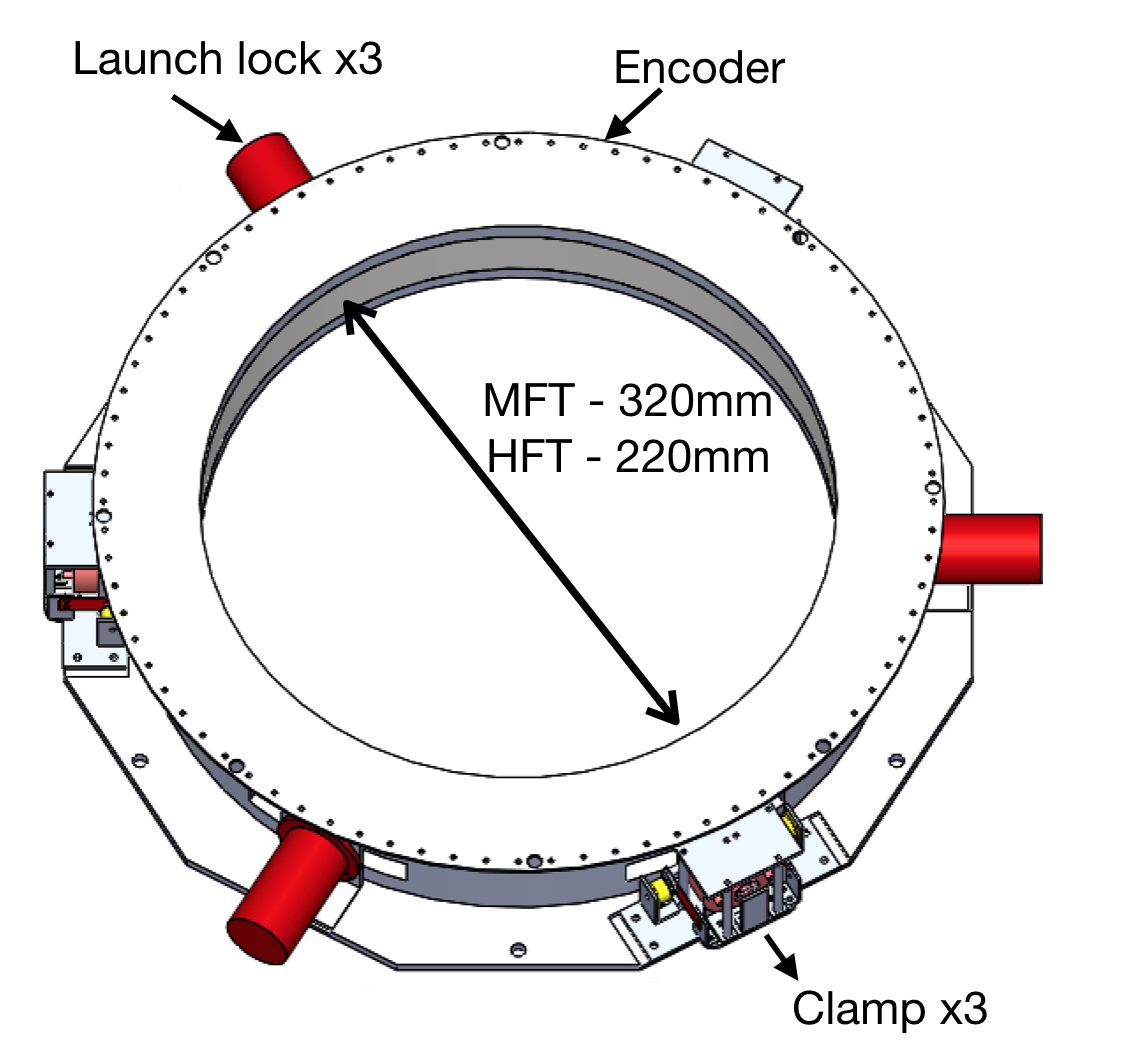}
   \end{tabular}
   \end{center}
   \caption[example] 
   { \label{fig:HWPM} 
Left: whole mechanism sketch. Right: section view from the 3D model.}
   \end{figure}

The main characteristics and requirements of the HWP rotating mechanism are a spin rate of 39\,rpm (61\,rpm), HWP diameter of 320\,mm (220\,mm) and angular accuracy ${<}\,1^\prime$ (${<}\,5^\prime$) for MFT (HFT).  The HWP temperature is ${<}\,20$\,K, the load on the 5-K stage ${<}\,4$mW and the total mass ${<}\,20\,$kg for both telescopes.
The cryogenic motor system is composed of some primary subsystems: the rotor assembly, which includes the SmCo ring magnet; the stator assembly, which includes the YBCO ring; the electromagnetic motor; and the encoder readout system.
The current rotator design is shown in Fig.~\ref{fig:HWPM} and is the same for both modulators, with a scaling of the components.

During the rocket launch, the rotor is held above the stator at room temperature by three pin pullers, radially oriented towards the center of the HWP ring. After the launch the pin pullers will be retracted and while the YBCO is cooling through its superconducting transition ($\sim \SI{90}{\kelvin}$), the rotor is held in position by an innovative frictionless clamp/release device \cite{Actuator:article}, based on electromagnetic actuators. 
This system will be used only once, but if needed it can clamp the rotor every time during the flight. 

The driver mechanism is conceptually similar to an electromagnetic motor: eight small SmCo magnets placed on the edge of the rotor are coupled with two rings of 32 coils each, on the top and bottom of the rotor, to obtain a larger and more uniform force. The current magnitude in each coil is adjusted using the feedback of the encoder readout system. The encoder consists of 64 equally spaced slits, in the periphery of the rotor, and LED emitters and photodiodes at room temperature, connected with optical fibers.
The expected total heat load of the driver system is $< \SI{4}{\milli\watt}$, where the dominant contribution comes from eddy currents and hysteresis losses. 

Simulations have been performed to estimate the thermal equilibrium of the rotor by simple radiative cooling. It has been demonstrated\cite{Columbro:2021} that operating the HWP at a maximum temperature of \SI{20}{\kelvin} could match both thermal equilibrium and sensitivity requirements.
The rotor temperature is also monitored by a custom capacitive sensors. The sensor is a thermistor, physically mounted on the rotating device and biased with an AC current, which is transferred from the steady electronics to the rotating device via capacitive coupling. A similar network of capacitors can monitor the levitation height.  The system reaches an accuracy better than 3\,\% for the measurement of the thermistor resistance, and $\sim\SI{10}{\micro\meter}$ for the measurement of its levitation height\cite{PdB_levitation_measurement2020}.

The issue of redundancy of this crucial mechanism has been addressed in more detail in Ref.~\citenum{Columbro:2021}. We did not perform any specific study for the pin puller harness material. We chose BeCu (Beryllium-Copper) wires (0.2--0.3\,mm thickness) as the standard for the modulator harness and this is also included in the PMU thermal budget\cite{Columbro:2021}. This choice minimizes both the conductive and the joule heat loads. This solution is also suitable for the pin-puller application (high fuse current in vacuum and very short current spike needed for the retraction).

\subsection{Detection chain}
\label{ss:detection-chain}


Both MFT and HFT focal-plane units (MF-FPU and HF-FPU) use superconducting TES bolometers\cite{Jaehnig:2020} cooled down to 100\,mK using a system based on seven adiabatic demagnetization refrigerator (ADR) stages that allow a 100\,\% duty cycle\cite{Duval:2020ADR}. At this base temperature, the detector arrays satisfy the instantaneous sensitivity required by the mission, including any noise added by detector readout electronics. The space-optimized DFMUX detector readout system is based on the technology used by the SPTpol, SPT-3G, POLARBEAR, and Simons Array ground-based instruments and flown on the EBEX balloon-borne experiment. 
The focal planes are composed of 3428 detectors (2074 for MFT and 1354 for HFT). The TES bolometers are coupled with silicon lenslets and sinuous antennas for MFT and coupled with silicon platelet feedhorns and orthomode-transducer (OMT) feeds for HFT. The focal planes include monochromatic, dichroic and trichroic pixels sensitive to polarization \cite{Suzuki:2018cuy}.
The LiteBIRD design is based on, and improves upon, the design currently used in the SPT-3G instrument, which has demonstrated 68$\times$ multiplexing with detector-limited noise performance \cite{Carter2018}. Two significant changes to the cold circuit are made from this design, while the warm electronics are radiation qualified and their redundancy improved. The bias element will be inductive and both the bias and the SQUID array amplifiers will be located on the 100-mK stage with the TES bolometers \cite{deHaan2020}.

\begin{figure} [ht]
   \begin{center}
   \begin{tabular}{c} 
   \includegraphics[width=0.9\textwidth]{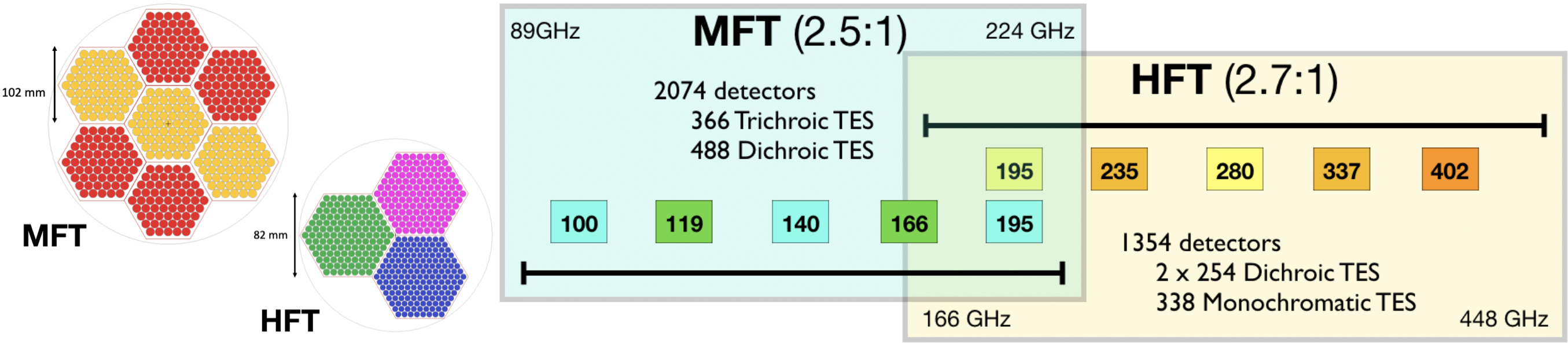}
   \end{tabular}
   \end{center}
   \caption[example] 
   { \label{fig:FPU} 
Left: schematic view of the focal detectors arrays. Right: frequency coverage and central frequency of the nine defined bands. It shows one overlapping band between MFT and HFT. The 2.5:1 and 2.7:1 labels give the bandwidths of the two optics, i.e., 224\,GHz/89\,GHz and 448\,GHz/166\,GHz, respectively.}
   \end{figure}

\begin{table}[ht]
\center
\begin{tabular}{cccccccc}
\hline
Telescope             & \begin{tabular}[c]{@{}c@{}}Detector\\ Type\end{tabular}                      & Module                 & \begin{tabular}[c]{@{}c@{}}Frequency\\ {[}GHz{]}\end{tabular} & \begin{tabular}[c]{@{}c@{}}Pixel Pitch \\ {[}mm{]}\end{tabular} & \begin{tabular}[c]{@{}c@{}}Module \\ Count\end{tabular} & \begin{tabular}[c]{@{}c@{}}Pixel \\ Count\end{tabular} & \begin{tabular}[c]{@{}c@{}}Detector\\ Count\end{tabular} \\ \hline
                      &                                                                              & MF1                    & {\pixelred{100/140/195}}                      & 12                                                             & 3                                                       & 183                                                    & 1098                                                     \\
 {\multirow{-2}{*}{MFT}} & {\multirow{-2}{*}{\begin{tabular}[c]{@{}c@{}}Lenslet/\\ Sinuous\end{tabular}}} & MF2   & {\pixelyellow{119/166}}                                & 12                                                             & 4                                                       & 244                                                    & 976                                                      \\ \hline
                      &                                                                              & HF1                    & {\pixelred{195/280}}                                & 7                                                              & 1                                                       & 127                                                    & 508                                                      \\
                      &                                                                              & HF2                    & {\pixelgreen{235/337}}                                & 7                                                              & 1                                                       & 127                                                    & 508                                                      \\
 {\multirow{-3}{*}{HFT}} &  {\multirow{-3}{*}{\begin{tabular}[c]{@{}c@{}}Horn/\\ OMT\end{tabular}}}        & HF3                    & {\pixelblue{402}}                                    & 6.1                                                            & 1                                                       & 169                                                    & 338 \\                             \hline                      
\end{tabular}
\caption{Focal plane configurations for the MH-FPU,  and HF-FPU. The colours of the frequencies correspond to those in Fig.~\ref{fig:FPU}.}
\label{tbl:focalplanes}
\end{table}


\subsection{Mechanical structure}
\label{ss:mechanical-structure}

The mechanical structure is composed of two main parts: the telescopes tubes and an exoskeleton. The telescopes tubes are designed to hold the optical elements and the various subsystems, and to ensure the optical alignment for each of the two telescopes. 
The exoskeleton holds the two tubes, ensures the required alignment, connects the telescopes to the payload module and provides the thermal link to the cryochain (Fig.~\ref{fig:Meca1}). All the mechanical elements are made of material compliant with the optical, mechanical, and thermal constraints, such as aluminum. To minimize the optical reflections, the internal surfaces of the tubes will be covered by an optical absorber.

Two options are still under study, the first one with a mechanical interface with the PLM on a mechanical ring linked to the 4.8-K stage of the cryo-chain, and a second option with a mechanical interface linked to the 30-K stage of the cryo-chain.
These two options are similar in terms of design of the telescopes, but the second one includes a cryo-mechanical structure between the 30-K stage and the 4.8-K stage to hold the MHFT.

\subsubsection{First option: mechanical interface at 4.8\,K}
\label{ss:mechanical-structure-base}

\begin{figure} [ht]
\begin{center}
\begin{tabular}{cc} 
\includegraphics[
 height=5cm]{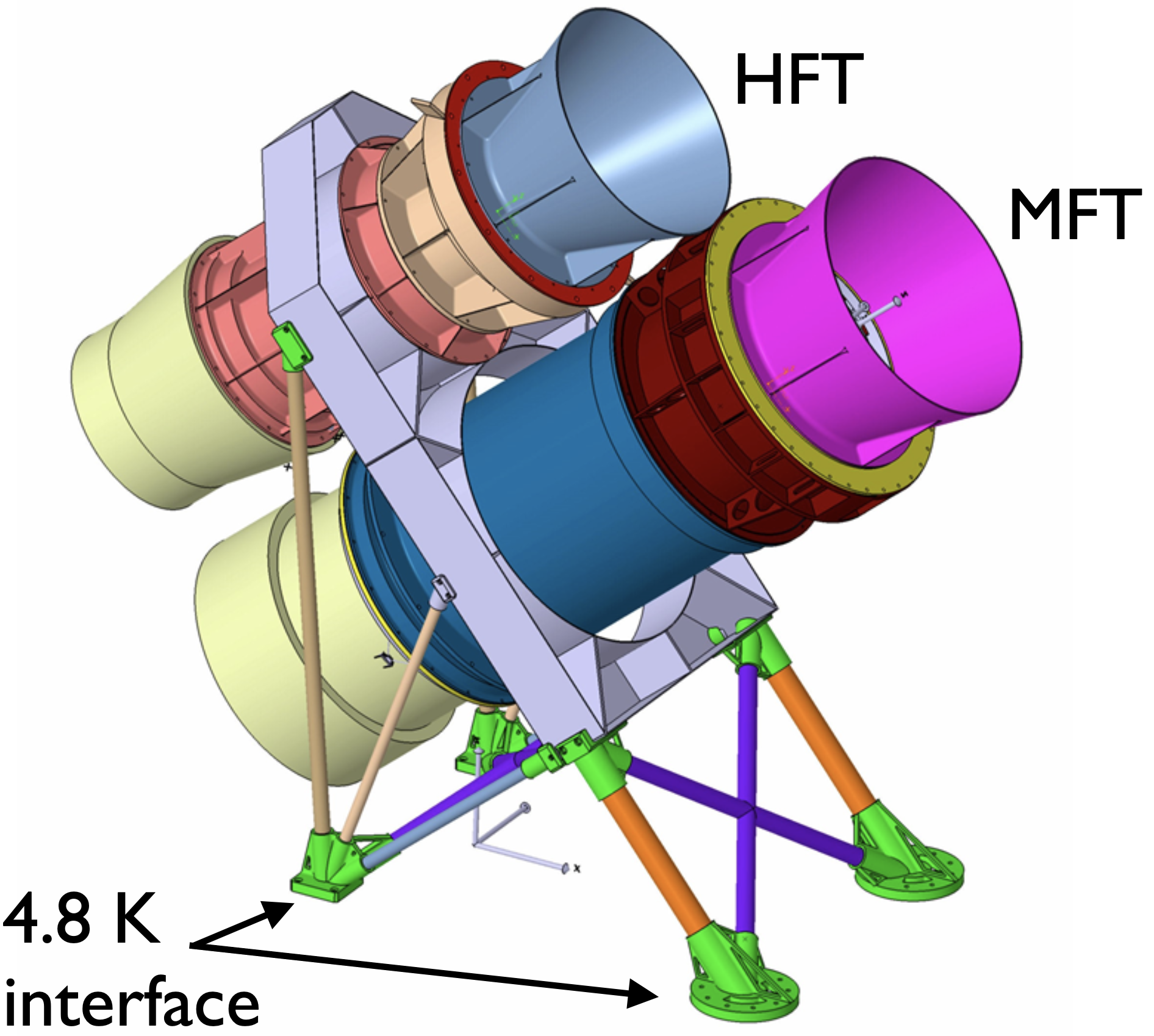}& 
  \includegraphics[
   height=5cm]{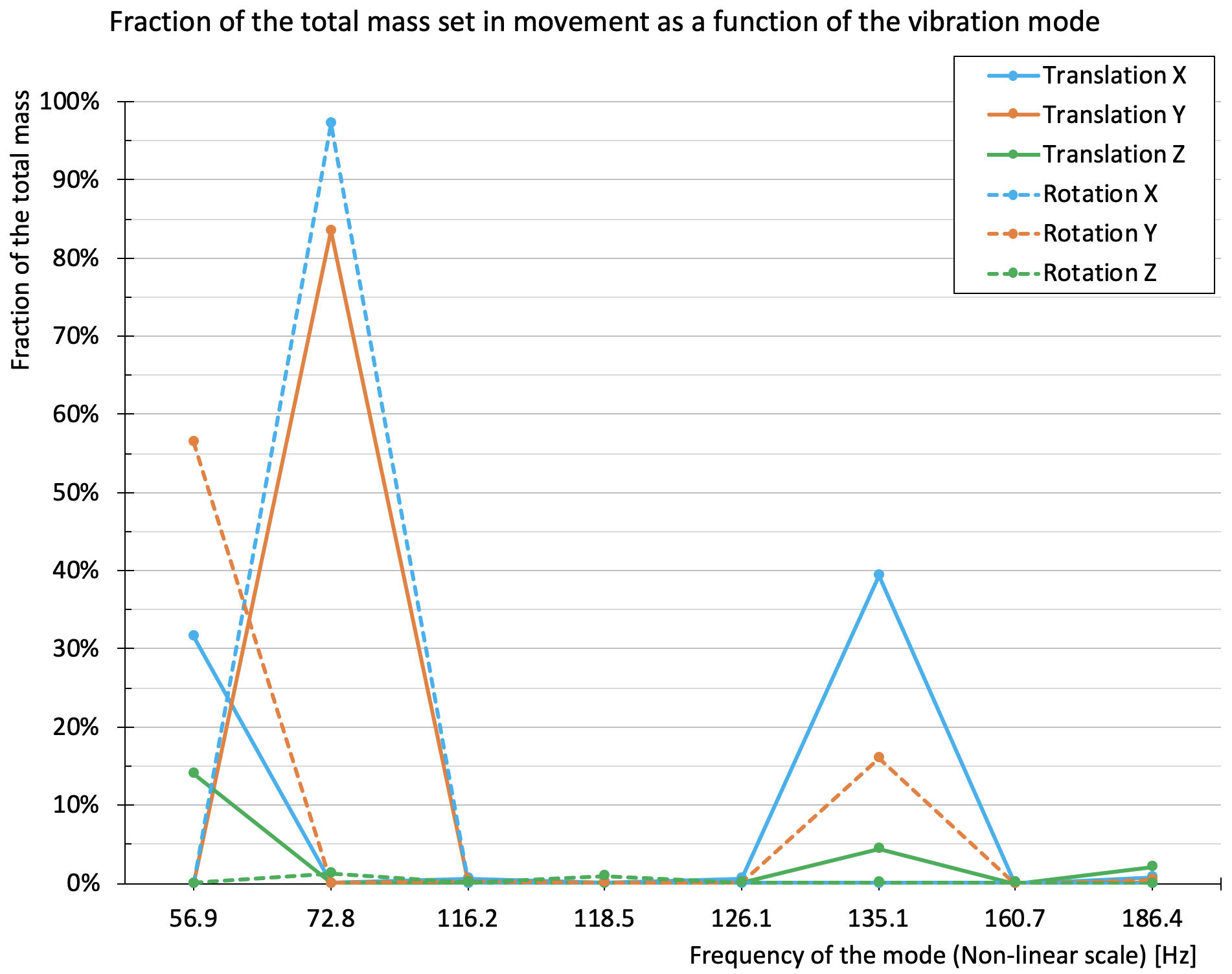}
\end{tabular}
\end{center}
\caption{Left: MHFT mechanical structure design based on a mechanical interface with the satellite on the 4.8-K stage. Right: Fraction of total mass study set in motion for each of the eight first modes.}
\label{fig:Meca1}
\end{figure}

A first series of iterations on the mechanical design have been performed to optimize the mechanical structure taking into account the various constraints such as minimum eigen-frequencies, total mass, launch load and thermal conduction. We finally converged on the design shown in the left panel of Fig.~\ref{fig:Meca1}. The MFT and the HFT tubes are held by the mechanical structure (plate and legs). They support all the MHFT subsystems and ensure the thermal link with the cryo-chain, the optical front baffles still have to be optimized with a stray-light analysis. The mechanical interface (ring at 4.8\,K) between the PLM and the MHFT is not shown here. While optimization is still ongoing, a first finite element modeling of the whole MHFT has been performed,
providing first estimates of the stiffness of the mechanical structure and the eigen-frequencies of the MHFT. 
The graph on right panel of Fig.~\ref{fig:Meca1} shows the fraction of the total mass set in movement as a function of the frequencies of the first modes. We consider here the modes with frequency under 200\,Hz, because modes larger than 200\,Hz are not relevant for our application. This study allows us to define which modes are significant (we only consider the modes impacting at least 5\,\% of the total mass) and gives us an estimate of the first structural modes.

This first analysis yields an estimate of the total mass of 118\,kg without margins, which is over the specification. Indeed, the requirement on the mass budget is to be lower than 100\,kg, including margins. This specification is driven by the parasitic heat load on the 4.8\,K stage due to the thermal conductance of the satellite mechanical structure holding both LFT and MHFT.
However this mechanical design reaches the goal in terms of eigen-frequency (the requirement on the first mode 100\,Hz [OZ] and 50\,Hz [OXY]), with no significant structural mode under 200\,Hz on [OZ] axis, and a first mode at 56\,Hz on [OXY] plane.

\begin{figure} [ht]
\begin{center}
\begin{tabular}{cc} 
   \includegraphics[
   height=5cm]{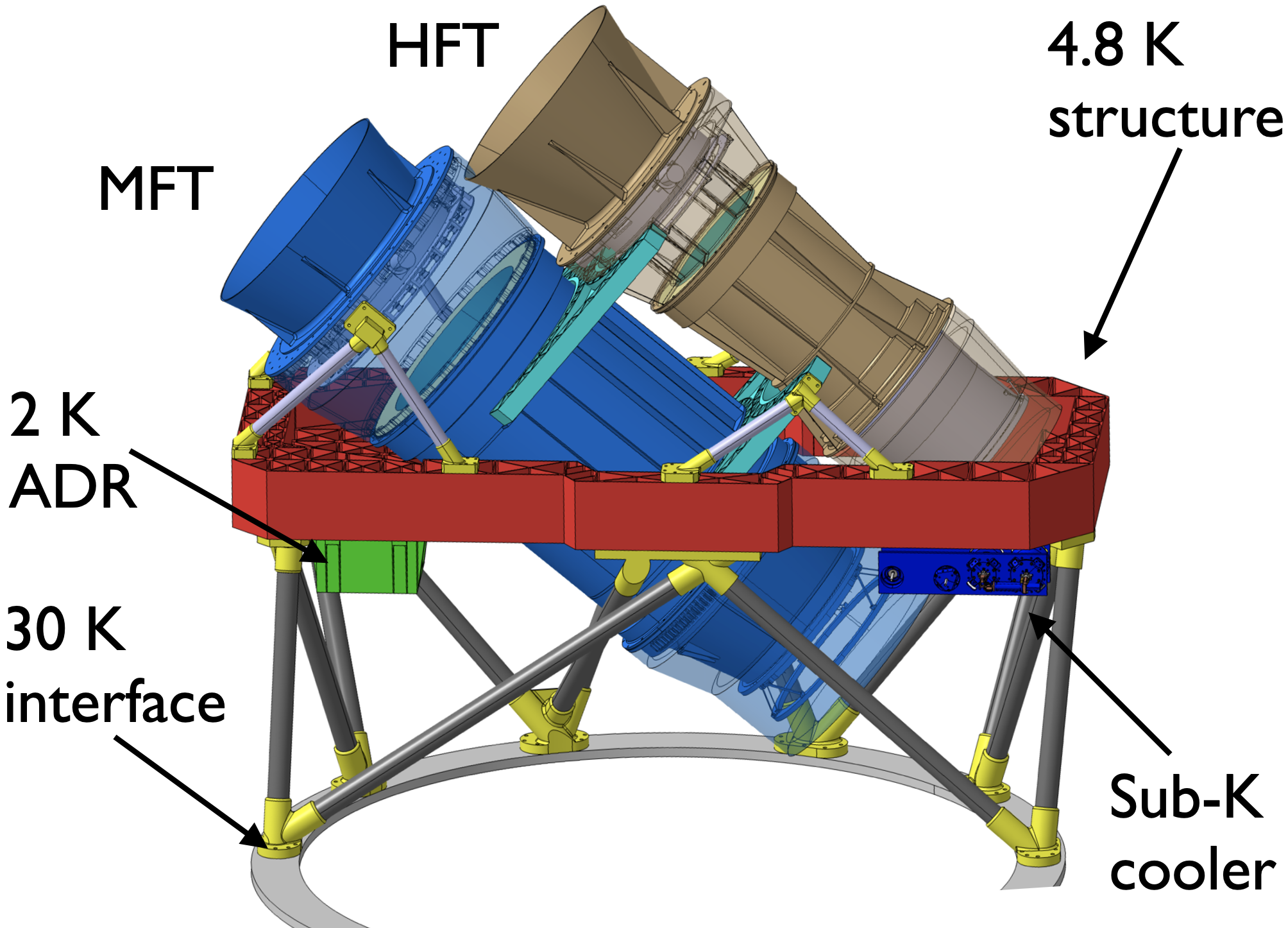} &
   \includegraphics[
    height=5cm]{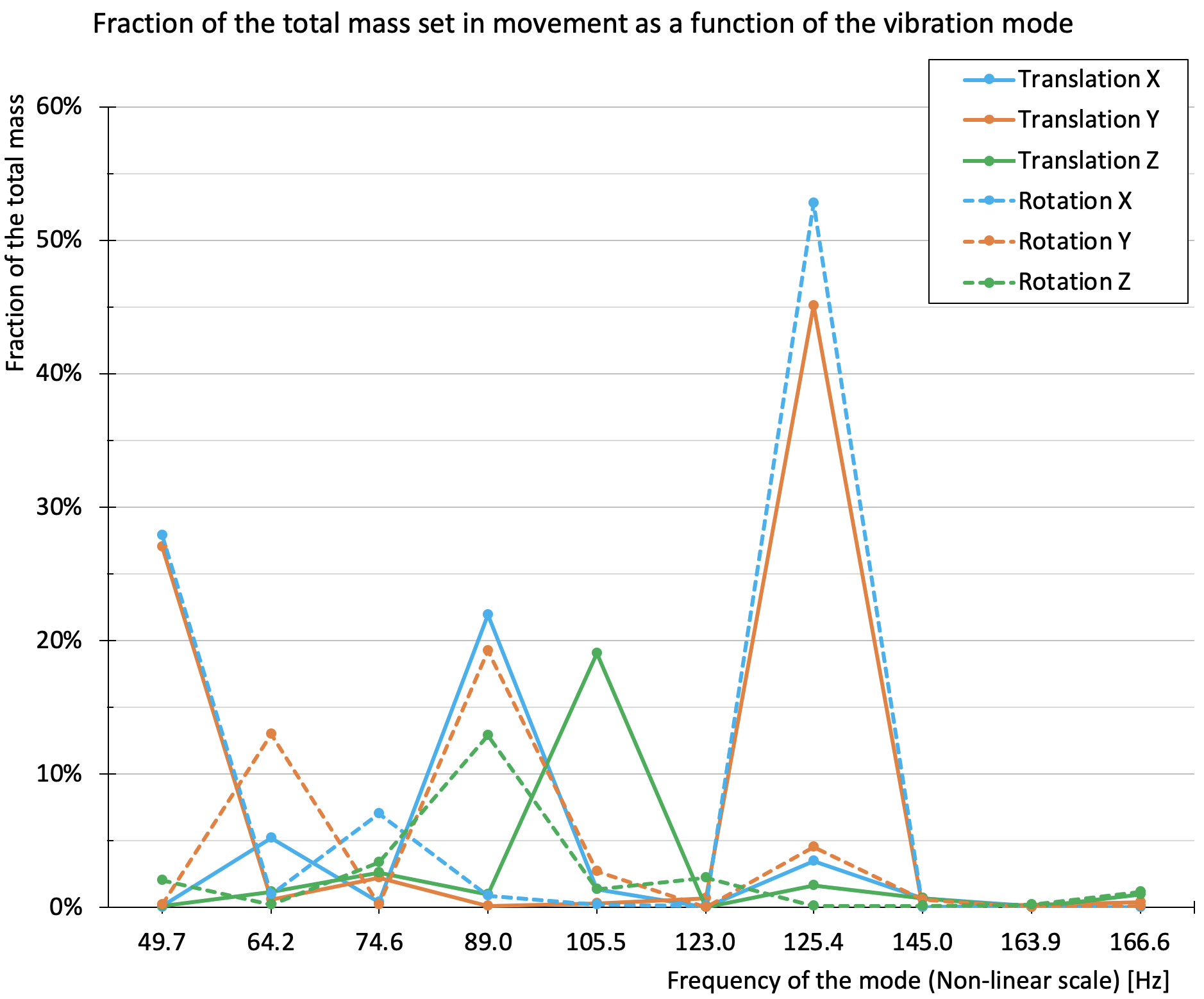}
\end{tabular}
\end{center}
\caption{Left: MHFT mechanical structure design based on a mechanical interface with the satellite at the 30-K stage. Right: Fraction of total mass study set in motion for each of the eight first modes.}
\label{fig:Meca2}
\end{figure}
   


\subsubsection{Second option: mechanical interface at 30\,K}
\label{ss:mechanical-structure-base}

To deal with the stringent constraints on the total mass budget at 4.8\,K, we recently decided to study a new option, which consists of moving the mechanical interface with the satellite from the 4.8-K stage to the 30-K stage of the cryochain.
 This new mechanical design is shown in the right panel of Fig.~\ref{fig:Meca2}. We designed the 30-K to 4.8-K cryomechanical structure with a particular attention to the thermal conductance. Indeed, the available cooling power on the 4.8-K stage is extremely tight (see section \ref{ss:challenge.thermal} for more details) and we need to limit as far as possible the parasitic heat load due to mechanical structure conductivity. Moving the interface from 4.8\,K to 30\,K allows us to gain mass on the 4.8-K ring interface and optimize the 4.8-K interface with the telescopes directly. In the same spirit, we assumed a cryomechanical structure made of CFRP (carbon-fiber-reinforced polymer) tubes to limit the parasitic heat load from the 30\,K stage while having a high stiffness. The requirements on the parasitic heat-load, (which should be smaller than 2\,mW), and the one on the eigen-frequency (first mode higher than 113\,Hz [OZ] and 53\,Hz [OXY]) are the two major drivers of the mechanical structure design.

   
While optimization has still to be performed, a first finite-element modeling of the whole MHFT, including the 30-K to 4.8-K cryomechanical structure, has been proposed, allowing us to get first estimates of stiffness and eigen-frequencies of the MHFT mechanical structure. We applied on this new design the same rules as for the first option, considering only the modes impacting at least 5\,\% of the total mass and with frequencies smaller than 200\,Hz. This FEM modeling provided us an estimate of the first structural mode. This initial analysis leads to an estimate of the total mass of 200\,kg including 25\,\% margins, which is higher than the mass estimate obtained for the first option described in \ref{ss:mechanical-structure-base}. The difference in the mass estimate, with the first option, is due the fact that more elements (like Cryo-coolers, thermal links, 30\,K to 4.8\,K cryo-mechanical structure,...) are know counted as MHFT subsystem.

The total mass budget has been revised in the framework of this second option. The heavy 4.8\,K interface ring, shared by LFT and MHFT in the first option design has now been replaced by two 30\,K to 4.8\,K optimized  cryo-mechanical structures. This solution allows us to drastically reduce the mass of the 4.8\,K interface with the satellite while minimizing the parasitic heat load on the coldest stages. It also implies a higher allowed mass budget for the instruments.

The constraints on the eigen-frequency and on the parasitic heat-load are satisfied. As shown in the right panel of Fig.~\ref{fig:Meca2}, the first structural mode is of 105.54\,Hz on the [OZ] axis (required at 100\,Hz) and of 49.67\,Hz on the [OXY] plane (required at 50\,Hz).
The estimated parasitic heat load due to the heat conductivity of the 30-K to 4.8-K cryomechanical structure is of about 1.9\,mW, which is within the specification of 2\,mW.




\section{MHFT challenges}
\label{s:challenge}

The achievement of the scientific breakthrough targeted by LiteBIRD requires associated technical advances to make it possible. We detail below the specific challenges we face when optimizing the design of the mid- and high-frequency telescopes, in the context of the global optimization of the LiteBIRD payload module.  

\subsection{Thermal}
\label{ss:challenge.thermal}


The design of the MHFT has
been mainly driven by two strong requirements: high sensitivity; and a very accurate control of the systematic effects. To reach these goals, a special effort has been put on the thermal control of the instruments, especially when dealing with the rotating mechanism of the HWP, the thermal stability of the mechanical structure, and the focal plane.

\begin{figure} [ht]
   \begin{center}
   \begin{tabular}{c} 
   \includegraphics[height=6cm]{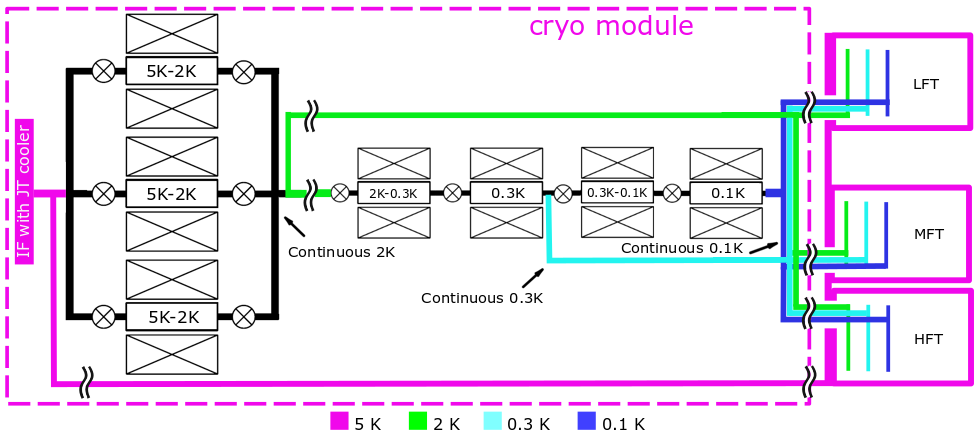}  
   
   \end{tabular}
   \end{center}
   \caption[example] 
   { \label{fig:MHFTcryo} 
LiteBIRD instrument cryomodule definition.}
   \end{figure}

\noindent
\textbf{\textit{A cryogenic instrument}}
In order to minimize the optical loading from the instrument itself on the focal plane, the full optics and mechanical structure of the MFT and HFT are cooled down to 4.8\,K. Furthermore, the optical hood, located between the secondary lens and the focal plane, and holding a set of filters, is cooled to 1.8\,K. This choice has multiple impacts on various aspects. The materials and their thermal-elastic properties have to be appropriate and well characterized at these cryogenic temperatures, as well as the emissivity properties of the optical elements, as detailed in sections \ref{ss:challenge.optical} and \ref{ss:challenge.calib}. Another key point is the limited resources of available cooling power required to cool down those instruments to 4.8\,K and below. The cooling power available for both MHFT and LFT is 17\,mW at 4.8\,K. This should cover the instrument dissipation and losses as well as the operation of the sub 4.8\,K cryomodule. Therefore the efficiency of the low temperature cooler is a key point. As presented in Fig.~\ref{fig:MHFTcryo}, the current design of the instrument cooling system\cite{Duval:2020ADR} is composed by a JAXA 4-K class JT cooler, a NASA 2-K cooler\cite{shiron:2018itomi}\cite{Shirron2012}, made of three ADR stages and a sub-K cooler from CNES-CEA, made of four ADR stages, providing continuous cooling power at 0.3\,K and 0.1\,K. The 2-K cooler and sub-K cooler will be integrated on the 4.8-K structure of the MHFT, as presented on Fig.~\ref{fig:Meca2}. Various thermal links will be implemented to connect the coolers together and cool down the instrument. A dedicated 2-K thermal link will connect the 2-K cooler to the instrument focal plans and will be used to intercept heat on the sub-K thermal link supports. In the context of the optimization of the design, a trade-off analysis between the cost in cooling power and the gain in sensitivity has been performed to decide on the temperature of the cold aperture stop (CAS) of the MFT and MHT, located just after the HWP mechanism. It has finally been shown that it is preferable to keep the CAS at 4.8\,K, reducing cooling power consumption without affecting too strongly the overall sensitivity.

\noindent
\textbf{\textit{Temperature stability}}
Even more important than the value of the temperature reference  for the various stages, the temperature stability of the instruments stages is crucial for the data analysis. The 0.1-K and 0.3-K heat sinks will be regulated at the sub-K cooler level at $1.0\,\mu{\rm K}/\sqrt{\rm Hz}$ on the frequency range of 0--1\,Hz. This will be possible except during the regenerative phases of the cold ADR stages. At these times the cold ADR buffers are facing quick thermal coupling and relatively high heat transfer. At 2\,K the approach is the same as for the colder stages, since the ADR operation implies switching phases, which disturbs the thermal control. At 4.8\,K the JT cooler is providing purely continuous cooling. Nevertheless, the JT cold tip is not thermal controlled as for the ADR. Thermal fluctuations of 10\,mK on a timescale of 10\,min are thus the base values for this cooler. Analyses are currently ongoing to obtain robust estimates of the specifications for temperature stability of all stages. 

In order to improve the thermal stability of this stage, the 2-K cooler-controlled heat-rejection process is foreseen to be used.
The other challenge is the thermal stability in the instruments themselves. Indeed, the liteBIRD instruments are sharing the same cryochain down to 0.1\,K and thermal links connect all the components to the cryomodule heat sinks (see Fig.~\ref{fig:MHFTcryo}). The ongoing dynamic thermal study of the MHFT will provide the performance thermal sensitivity parameters useful in the definition of the sub-system requirements, as well as the cryo-design and active thermal control needs.

These stringent thermal stability requirements ultimately impact the processing of the levitating HWP. It was initially envisaged to keep the rotating HWP at 4.8\,K by clamping it to cool it down on a daily basis after an inevitable increase of its temperature because of eddy-currents inside the supra-conductor material. Such a process would have induced large temperature variations of the HWP, which would have had a dramatic impact on the stability of the heat load of the FPUs. It has been finally chosen to cool down radiatively (and continuously) the HWP rotors and to maintain an equilibrium temperature around 18\,K.

\subsection{Optical}
\label{ss:challenge.optical}


The optical modeling of the MHFT is an ongoing effort that relates to both sensitivity calculations and overall performance evaluation from full-sky time-domain simulations. Optical systematics can be caused by various non-idealities along the optical chain, including lenses, forebaffle, rotating HWP, focal plane baffle, filters, etc. Ideally, all such issues need to be accounted for, both at the system and subsystem level, to inform design drivers and possibly feed real data into simulations for validation purposes in view of systematics studies.
Here we briefly highlight optical components (and related physical effects) that require particular care and improvement with respect to the present level of understanding. 

\noindent
\textbf{\textit{Metal-mesh components}}
Similarly to the HWP, optical filtering through the cryogenic chain is achieved by multi-layered metal mesh interference filters\cite{Ade2006_SPIE}.  These filters have been deployed for many years on ground-based, balloon-borne and satellite instruments such as Planck\cite{PlanckHFI_optical_design} and Herschel\cite{Herschel}) including CMB polarimetry experiments (BICEP/KECK\cite{bicep2keck}, ACTpol\cite{AdvancedActFieldedPerformance}, EBEX\cite{ebex}, SPIDER\cite{SPIDER}, POLARBEAR\cite{Polarbear:article}, etc.) and are therefore technically mature and well qualified.  The filters are manufactured from polypropylene and copper; typically consisting of between 6 and 16 layers of photolithographically etched copper, which are precisely aligned and dielectrically spaced, then hot-bonded together. The devices predominantly reflect unwanted radiation and will reach thermal equilibrium with their radiation environment. These devices can be fabricated in diameters up to 700\,mm.

In the case of LiteBIRD, the challenge comes from optimizing thermal filtering strategies to take advantage of the low thermal background afforded to space-borne observations with access to limited cooling power. Placement and equilibrium temperatures of 40-cm diameter metal mesh filters can critically impact both cryogenic performance and detector loading, especially at high frequencies. This will require significant thermal modeling effort and pre-flight testing. 

\noindent
\textbf{\textit{Instrumental polarization}}
Imperfections in the HWP yield a plethora of known effects on the polarized signal modulation.
These most likely appear in time-ordered data for each detector and peak at harmonics of the modulation frequency in the signal frequency spectrum. 
Many such effects have been modeled and extensively studied in CMB observations from balloon-borne and ground-based experiments, and techniques exist to characterize and remove them from recorded signals at the data analysis level\cite{ritacco2017}. 
In addition, any departure of the beam from a Gaussian shape can cause leakage of the total intensity into polarization, resulting in artifacts in the co-polar beam pattern \cite{ritacco2017}. This effect might be observed in the Stokes $Q$ and $U$ maps of an unpolarized source and is generally referred to as instrumental polarization. Its modeling is usually more difficult \cite{2020EPJWC.22800002A} for large-format arrays because each detector might have a slightly different beam. For this reason the knowledge of the telescope beams, the modeling and characterization of detector coupling at each frequency, and the simultaneous study of these effects through simulations and laboratory tests with the full optics, including the HWP, is crucial to constrain systematic effects arising in the final data.

\noindent
\textbf{\textit{Ghosts}}
Optical ghosting arises whenever light gets reflected or scattered off surfaces due to imperfect impedance matching, e.g., on the filters and lens surfaces in the presence of non-ideal ARCs, at the focal plane level as a consequence of return loss of the feeding elements (lenslets or feedhorns), or on the tube walls as a consequence of imperfect or angle-dependent absorption by absorber coatings. If this light undergoes an even number of such scatterings it can be redistributed, or even partially refocused, on the focal plane, resulting in an uneven detector power distribution and possibly image artifacts. A large number of different non-idealities and interactions may combine into unwanted effects, making it difficult to reproduce through a pure modeling/simulation effort. Some effects, like polarization-sensitive scattering are intrinsically difficult to code into a robust simulation environment and a suitable experimental characterization is mandated to validate any modeling effort in this sense.

\noindent
\textbf{\textit{Absorbers}}
The use of absorbing coatings and materials,
complemented by careful shaping of the internal surfaces of the tube assemblies, is a standard technique to mitigate issues with stray radiation, in association to a proper spectral rejection performed at the level of the filtering chain. 
Ideally, an absorber emits a perfect blackbody spectrum (in which case its contribution to the power loading on the focal planes depends only on the absorber temperature), efficiently absorbs radiation well away from normal incidence, and features a large thermal inertia, so that the radiating/absorbing properties do not exhibit thermal drifts that might ultimately determine dangerous time-constant effects in the detectors response. In addition, any candidate absorber for the MHFT tubes must comply with the mass allocation constraints and meet the requirements for space qualification.
MHFT absorbers need to be carefully measured and
modeled to assess their behavior at low temperature (the MFT and HFT tubes operate at 5\,K) in terms of
incidence-, polarization- and frequency-dependent reflectance and scattering. Most of these effects
cannot be easily implemented in numerical simulations and their treatment must rely on dedicated experimental tests to inform system-level modeling.

\subsection{Detection chain}
\label{ss:challenge.detection-chain}


\noindent
\textbf{\textit{Multi-chroic space optimized detectors.}} The two detector technologies chosen for LiteBIRD have been successfully deployed on many ground-based \cite{SPT3G2year, PB2AndSATokiLTD, AdvancedActFieldedPerformance} and balloon-borne \cite{Spider280GHz} mm and sub-mm astronomy experiments.  More details on the detection chain can be found in a companion SPIE proceedings on the detector fabrication status \cite{Westbrook2020}.  In order to meet the sensitivity requirements of the mission, several key advancements to these key technologies are required. In particular the ambitious frequency coverage of LiteBIRD, requires the development of many unique types of filter.   SPT-3G has successfully deployed a trichroic focal plane with bands centred at 90/150/220\,GHz, which is similar to the MF1 modules. However the filters do require specific tailoring of the circuitry to meet the bandwidth designs for the entire experiment.  Both NIST and UCB have fabricated triplexing filters appropriate for the LFT and will have a very similar design to both type of MF-FPMs. The bands must be accurate to within 2\,\% of their central frequency and have very low cross-talk between different pixel types. 

The HFT-FPUs are most similar to the 280-GHz focal plane units fabricated by NIST for the second flight of the SPIDER experiment (Spider~2) \cite{Spider280GHz}.  These arrays were monochromatic 280-GHz feed-horn-coupled OMT arrays fabricated with excellent yield and cross-wafer uniformity.   The design team at NIST has fabricated dichroic HF1 prototype pixels and characterization is under way. 

The LiteBIRD mission also has the lowest instantaneous sensitivity ($NEP$) of any CMB polarization experiment.  The bolometers for each band have thermal conductance levels tuned such that the operating power is 2--3 times larger than the optical power on the bolometer.  Generally speaking the bands for the MHFT will have an optical load of 0.2 to $0.5\,\mathrm{pW}$ and therefore will have saturation powers between 0.4 and $1.5\,\mathrm{pW}$. To achieve the $NET$ requirement once the $NEP$ requirements are met the focal plane arrays must have tip-of-the feed to bolometer island optical efficiency of $\geq$ 60\,\%, which has been demonstrated on many ground-based and balloon-borne experiments \cite{Carter2018, Spider280GHz, WestbrookSimonsArrayFabrication, AdvancedActFieldedPerformance}.

\noindent
\textbf{\textit{Frequency-domain multiplexing. }} The MHFT telescope will operate detectors using a digital frequency-domain multiplexing (DfMUX) \cite{Dobbs2008} readout system based on the design currently in use on SPT-3G. \cite{Bender2016} This technology allows multiple detectors to be operated with a single set of electronics and cryogenic superconducting quantum interference device (SQUID) amplifiers. TES readout technology is characterized by: the number of detectors that can be operated with a single multiplexing module (the \textit{mux factor}); the fractional increase in total noise due to the readout; and the crosstalk fraction between detector pairs within a multiplexing module. For a space-based platform such as LiteBIRD there are additional constraints on mass, power consumption, and reliability. The LiteBIRD DfMUX design will have the same mux factor as existing ground-based instruments, but more strict design requirements for readout noise and crosstalk. Readout noise must be subdominant to the detector noise, and LiteBIRD detectors are significantly lower noise than equivalent ground-based detectors because they avoid radiative atmospheric loading. Consequently, the acceptable readout noise for the LiteBIRD DfMUX system is lower than for a ground-based system. Similarly, the increased sensitivity of the LiteBIRD system dictates a factor-of-several improvement in crosstalk relative to existing ground-based systems such as SPT-3G. A space-flight qualified DfMUX system is being designed by the LiteBIRD Canada group as part a technology development program funded by the Canadian Space Agency, begun in 2012.\cite{Bender2014, Montgomery2015} A demonstration model of the LiteBIRD readout will be ready in 2021, and preliminary forecasting indicates the baseline configuration will meet the above requirements.\cite{Montgomery2020}

\noindent
\textbf{\textit{Cosmic rays.}}
LiteBIRD will observe the CMB for 3 years from the second Earth-sun Lagrange point (L2). The LiteBIRD mission will be operational during a period of minimal solar activity, during which time the flux of Galactic cosmic rays (GCRs) will be maximized. In order to predict these effects, we have produced an end-to-end simulator for cosmic ray effects in LiteBIRD \cite{Stever2020,Tominaga2020}, which provides an estimate for the level of degradation to mission success from cosmic ray systematic effects, we well as a tool to evaluate sensitivity and potential design changes. Our preliminary simulations have found that the cosmic-ray signal is similar to white noise, owing to the very large number of hits per second into the geometrical area of one detector wafer, and the common thermal mass between all of the detectors in a given wafer. 

The LiteBIRD detector pixels are designed with on-chip physical mitigation mechanisms for avoiding the propagation of athermal phonons over large areas of the detector wafer. The majority of cosmic-ray impacts will occur into the detector wafer rather than into the TES bolometers themselves, owing to the small size of the TESs. The summed thermal fluctuation across the wafer from many subsequent energy depositions is expected to be the main source of cosmic-ray noise, rather than large direct hits as was the case for Planck-HFI. It is for this reason that the LiteBIRD pixel design aims to include both mitigation mechanisms against athermal phonons and strong thermal dissipation to the invar frame. 


\noindent
\textbf{\textit{Data rate.}} 
The number of detector channels has increased by two orders from the Planck satellite. The telemetry rate between the spacecraft and the ground stations does not increase proportionally. This means that we need to reduce the data rate for each detector in comparison to Planck. In Planck-HFI, a data rate of 180~Hz per detector was achieved. In LiteBIRD, in contrast, the data rate per detector is 20\,Hz in the current design. Information higher than 10\,Hz is not only lost but also contaminates the 10\,Hz bandpass by aliasing. We remove aliasing by a combination of an anti-aliasing analog filter and multi-stage low-pass digital filters. Because the HWP modulates the target signal at 2--4\,Hz, the 10-Hz band-pass is sufficient to achieve our science goals. This, however, limits our ability to diagnose fast signals such as cosmic-ray glitches. In comparison to Planck-HFI, LiteBIRD has a slower data rate and a faster detector time constant due to the thermoelectric feedback of the TESs. We would not be able to resolve individual glitches in the time domain and the deglitching technique developed for Planck-HFI \cite{2014A&A...571A..10P} will not be employed. We plan to remove cosmic-ray power in the frequency domain,
which is described in Refs.~\citenum{Stever2020} and \citenum{Tominaga2020} in this volume.

\subsection{Electromagnetic interference}
\label{ss:challenge.emi-emc}


Low-temperature detectors are sensitive to any form of energy input that thermalizes. We
also require a very low noise level of a few aW/$\sqrt{\mathrm{Hz}}$ level in a spacecraft that is densely packed with electronics consuming a total of 3\,kW. Electromagnetic interference to the detector and the readout system by external components is a serious challenge. In fact, previous cryogenic missions observed significant, though manageable levels of EMI, Planck-HFI by line noise caused by the cryocooler drive frequency, and Hitomi-SXS by line noise through the pulse-width modulation frequency of the magnetic torquer. Both of these two missions employed high-impedance sensors unlike the TESs to be used in LiteBIRD, thus their experience would not be completely applicable. Nevertheless, it is important to identify the EMI-coupling routes and decrease their risk level as early as possible in the mission.

The high-risk EMI items include: (a) magnetic interference to the TES detectors and the
SQUID amplifiers; (b) high-frequency EMI interference from the ground-link communication
to the focal plane instruments; and (c) conductive EMI from high-power consuming
components through bus voltage ripples penetrating into the focal plane. For (b), we started a simulation study based on EMI modeling and the
initial results are presented in \cite{tsuji2020} in this volume. For (c), we plan to perform
simulations as well as component-level tests at an early stage of the project by following the heritage of Planck-HFI.

\noindent
\textbf{\textit{TES Detectors and SQUID amplifier suceptibility to EMI.}} An additional challenge of the detection chain for LiteBIRD is the very stringent noise performance requirements.  LiteBIRD requires that all of the detectors have an unmodulated $1/f$-knee of $20\,\mathrm{mHz}$, which requires careful attention to the focal-plane structure.  We anticipate that $1/f$ noise arising from variations in magnetic fields, thermal drifts of the focal plane, micro-vibrations coupling to the TESs, electromagnetic interference, and cosmic rays interfering with the operation of the focal planes. 
The focal-plane structure will create a Faraday cage around the SQUID electronics at 100\,mK to help shield the SQUIDs from any EMI present inside the cryostat. 

\noindent
\textbf{\textit{ADRs magnetic field and EMI.}}
The coolers have relatively high operating magnetic fields (0.6 to 2\,Tesla), which are passively shielded. A remaining field of 200\,$\mu$T at 100\,mm from the cooler's outer shell, is taken as the basic value for the sensitivity study. As presented in Fig.~\ref{fig:Meca2}, in order to minimize the magnetic field at the detector level, the 2-K cooler has been located as far as possible on the 4.8-K MHFT structure. The sub-K cooler, on the contrary, has been kept close to the focal-plane interfaces to reduce thermal gradients at these sensitive temperature levels. This is possible because the magnetic fields are lower for this cooler. 
In terms of dynamic aspects, the ADR stages are operating periodically between slow magnetic field ramping during isothermal heat exchanges at warm and cold temperatures and fast high magnetic field ramping during adiabatic temperature variations between the warm and cold temperatures. The main fluctuation seen by the focal planes are thus the latter and we consider a 0 to 200\,$\mu$T variation in less than 5\,minutes. 
Part of the design study on the ADR cooler will be to optimize the location and orientation of the ADR coils in order to minimize the remaining magnetic field at the detector level. The questions about radiative and conductive EMI will also be addressed during the thermal links design phase and cooler Faraday cage definition.

\noindent
\textbf{\textit{PMUs susceptibility/impacts.}}
Both polarization modulator units have a variable magnetic field. The magnitude of the magnetic field close to magnetic levitation is $\sim{0.5}\,$T and its inhomogeneities are of the order of 1\,\%. This field is passively shielded by a magnetic shield in the radial direction to minimize the interaction between PMUs and with the LFT focal plane. On the other hand, due to the on-axis design of the MHFT, the magnetic field cannot be shielded in the axial direction. Thanks to the distance between the MHFT focal planes and the relative PMUs, the residual magnetic field on the detectors is ${<}\,100\mu$T with a variation ${<}\,1\mu$T, synchronous with the HWP rotation frequency.

\subsection{Calibrations}
\label{ss:challenge.calib}

The calibration plans of MHFT are defined together with those of LFT, considering LiteBIRD as a single unified instrument. Given the stringent requirements defined in Sect.~\ref{ss:requirements}, we can define the two main challenges as being the calibrations of the beams and the spectro-polarimetric properties. For both types of parameters, the best strategy is to combine components, sub-systems and integrated measurements, on the ground and in-flight, together with precise modelling.  

For the beam characterization, two main requirements are considered: we need to specify the beams down to $-$56\,dB with respect to the main beam amplitude; and the regime between $-$20 and $-$35 \,dB has to be determined to better than 10\,\%. To reach these levels, the co-pol main and near sidelobe beams are planned to be reconstructed from planet observations, while the cross-pol response and the far sidelobes will be characterized by a combination of ground measurements and optical modelling, at warm and cold temperature in a cryogenic compact-antenna test range (CATR). To consolidate our ability to accurately model and measure the RF properties in the next 2 years, we plan to develop a bread-board model, which consists of a one- and two-lens system that will be modeled and characterized in 2020. In parallel we will study the feasibility of a cryogenic CATR by developing dedicated (large-scale) absorbers. Finally a prototype of one of the MHFT (opto-mechanical) tubes will be designed and fabricated, to be later characterized in a dedicated warm CATR. 

For the spectro-polarimetric properties, dedicated studies showed that the spectral resolution should be $\Delta \nu \simeq 0.2$--2\,GHz, depending on the frequencies\cite{Ghigna:2020wat}, while the knowledge of the polarization angle should reach a few arcminutes
for the (more sensitive) CMB channels. The plans are to calibrate
these parameters on the ground (for each integrated instrument independently)
within a cold flight-like environment, with a special emphasis on
the relative polarization angle. This should be performed with a dedicated
ground segment equipped with two orthogonal inputs (FTS, VNA or another coherent source).
A measurement of the spectra at each small variation of HWP angle position will then be
required in order to reconstruct the expected modulation curve for any given spectral index source.\cite{Savini2009}
This will be complemented with in-flight data analysis. The Crab Nebula is foreseen being used,
even though a dedicated and coordinated ground measurement campaign will be required
to meet the required accuracy on the knowledge of its polarization angle. Ongoing studies
based on previous publications\cite{{Aumont2010},{weiland/etal:2018},{pccs2planck},{Ritacco}} 
show that the expected LiteBIRD sensitivity 
should allow us to reach a few arcminutes accuracy in each band. Nulling
the $EB$ cross-correlation is also foreseen.
It has been shown\cite{MinamiKomatsu:2020} that we can calibrate LiteBIRD detectors with uncertainties 
of the order of $2.7$\,arcmin, even though this would require to give up measuring the birefringence.






\section{Concluding remarks}
\label{s:conclusion}

The LiteBIRD mission\cite{Hazumi2020} has been selected in 2019 by ISAS/JAXA as a Large-Class mission for a launch in 2029. The main international partners have already entered in Phase~A, in order to perform the optimization of the instrumental design, to increase performance of the key components and to demonstrate the overall feasibility of the concept. 
We have reported here the current status of the design of the mid- and high-frequency telescopes, so-called MFT and HFT, which are under European responsibility, but including also major  international contributions, such as focal planes or warm-readout electronics. 
We stress that the performance of the MFT and HFT telescopes should be combined with those of the low-frequency telescope\cite{Sekimoto2020} (LFT) to assess the overall performance of the LiteBIRD mission, and cannot be considered separately, since a global optimization of the three telescopes is required to achieve the expected sensitivity on the tensor-to-scalar ratio of $\delta(r) < 0.001$. We have emphasized that such a challenging scientific goal requires facing technical challenges. While a lot of effort has already been injected into the development of new technologies for LiteBIRD applications, and deep analyses are currently being performed to assess the detailed technical requirements of MFT and HFT, we still have to deal with challenging issues.  This includes the adaptation to space conditions of the high-sensitivity and low-noise detection chain, the thermal control and modeling of this cryogenic instrument, the development of the magnetically levitating continuously rotating HWP, the optical modeling of various quasi-optical components of these refractive instruments, and also the calibration campaign in a flight-like environment. In this context, we expect important progress to be achieved in the coming years.

\appendix    

\acknowledgments 
 
{\small  
This work is supported in Japan by ISAS/JAXA for Pre-Phase A2 studies, by the acceleration program of JAXA research and development directorate, by the World Premier International Research Center Initiative (WPI) of MEXT, by the JSPS Core-to-Core Program of A. Advanced Research Networks, and by JSPS KAKENHI Grant Numbers JP15H05891, JP17H01115, and JP17H01125. The Italian LiteBIRD phase A contribution is supported by the Italian Space Agency (ASI Grants No. 2020-9-HH.0 and 2016-24-H.1-2018), the National Institute for Nuclear Physics (INFN) and the National Institute for Astrophysics (INAF). The French LiteBIRD phase A contribution is supported by the Centre National d’Etudes Spatiale (CNES), by the Centre National de la Recherche Scientifique (CNRS), and by the Commissariat à l’Energie Atomique (CEA). The Canadian contribution is supported by the Canadian Space Agency. The US contribution is supported by NASA grant no. 80NSSC18K0132. 
Norwegian participation in LiteBIRD is supported by the Research Council of Norway (Grant No. 263011). The Spanish LiteBIRD phase A contribution is supported by the Spanish Agencia Estatal de Investigación (AEI), project refs. PID2019-110610RB-C21 and AYA2017-84185-P. Funds that support the Swedish contributions come from the Swedish National Space Agency (SNSA/Rymdstyrelsen) and the Swedish Research Council (Reg. no. 2019-03959). The German participation in LiteBIRD is supported in part by the Excellence Cluster ORIGINS, which is funded by the Deutsche Forschungsgemeinschaft (DFG, German Research Foundation) under Germany’s Excellence Strategy (Grant No. EXC-2094 - 390783311). This research used resources of the Central Computing System owned and operated by the Computing Research Center at KEK, as well as resources of the National Energy Research Scientific Computing Center, a DOE Office of Science User Facility supported by the Office of Science of the U.S. Department of Energy. European collaborators acknowledge support from the European Research Council (ERC) under the European Union’s Horizon 2020 research and innovation programme (grant agreement Nos. 772253, 819478, and 849169). The European Space Agency (ESA) has led a Concurrent Design Facility study, focused on the MHFT and Sub-Kelvin coolers, and funded Technology Research Programmes for “Large radii Half-Wave Plate (HWP) development” (contract number: 4000123266/18/NL/AF) and for the ‘Development of Large Anti‐Reflection Coated Lenses for Passive (Sub)Millimeter‐Wave Science Instruments” (contract number: 4000128517/19/NL/AS).
}



\end{document}